\documentclass{raa}            

\usepackage{graphicx,times}             

\begin{document}

   \title{Simulation of Ultra-Long Wavelength interferometer in  the Earth orbit and on the lunar surface
}

   \volnopage{Vol.0 (200x) No.0, 000--000}      
   \setcounter{page}{1}          

   \author{Mo Zhang
      \inst{}
   \and Mao-Hai Huang
      \inst{}
   \and Yi-Hua Yan
      \inst{}
   }

   \institute{National Astronomical Observatories, Chinese Academy of Sciences,
             Beijing 100012, China; {\it mhuang@nao.cas.cn}
   }

   \date{Received~~2009 month day; accepted~~2009~~month day}

\abstract{We present simulations for interferometer in the Earth orbit and on the lunar surface to guide the design and optimization of space-based Ultra-Long Wavelength missions, such as those of China's Chang'E Program. We choose parameters and present simulations using simulated data to identify inter-dependencies and constraints on science and engineering parameters. A regolith model is created for the lunar surface array simulation, the results show that the lunar regolith will have an undesirable effect on the observation. We estimate data transmission requirement, calculate sensitivities for both cases, and discuss the trade-off between brightness temperature sensitivity and angular resolution for the Earth orbit array case.
}

   \authorrunning{M. Zhang et al.}            
   \titlerunning{Simulation of ULW Interferometer  in the Earth Orbit and on the Lunar Surface}  

   \maketitle

%
%
\section{Introduction}           
\label{sect:intro}

The Ultra-Long Wavelength (ULW)  spectrum window in astrophysics is generally defined as the radio band at wavelengths longer than 10 m. In this paper, we emphasize wavelengths between 10\,m and 10\,km (0.03$\sim$30\,MHz). The ULW is a potentially significant frequency range for radio astronomy but currently has not been well explored (Jester \& Falcke~\cite{JF09}). Because of observational constraints from the ionosphere of the Earth, man-made and natural radio frequency interferences (RFIs), astronomical observations in this frequency range are more readily conducted from the space. Antennas deployed in the Earth orbit or on the lunar surface (Jester \& Falcke~\cite{JF09}) are able to receive signals well below 10\,MHz, being mainly limited by the interplanetary plasma cutoff frequency of 20$\sim$30\,kHz. Compared with ground-based observations, space interferometer allows for baselines many times longer than the diameter of the Earth, making much higher angular resolutions achievable even at ULW bands.

For astronomical space interferometry observations, requirements on array configuration and observation time windows vary greatly, depending on the properties of the celestial objects and the nature of the astronomical phenomena to be observed. Observations of compact sources require high angular resolution, therefore long baselines are necessary, while studies of faint extended sources requires high surface brightness sensitivity, which demands high array filling factors and therefore shorter baselines are essential. A full-sky survey mission would also impose different requirements compared with a time-domain event monitor.

Science requirements, engineering constraints and environmental obstacles must all be evaluated quantitatively in a trade-off study. Given the highly different, and sometimes contradictory, requirements of astronomical space interferometry missions, the mission outcome needs to be evaluated quantitatively against a set of science and engineering requirements and environmental constraints so that trade-offs appropriate to a space ULW mission can be made. Engineering factors to be considered include the accuracy of baseline determination, time synchronization, onboard processing and data transmission capabilities, and, for a lunar surface mission, power supply and antenna deployment. Environmental effects include the effects of space plasma and radio interference. Interplanetary, interstellar, and lunar plasma degrade observation results through angular broadening, time-delay, and Faraday rotation effects. ULW terrestrial natural and man-made radio interference that escapes from the Earth's ionosphere are among the main obstacles to overcome for ULW observations conducted near the Earth (Woan~\cite{Woan2000}). 

This paper aims to establish a framework for the trade-off study. Specifically, we simulate observations made by an antenna array for two cases: an array deployed in the Earth orbit, and an array deployed on the lunar surface. We first select the orbital parameters for the Earth orbit case and antenna positions for the lunar surface case and then obtain the $(u,v)$-coverage for each, then we simulate the signal that the antennas receive and use these to create images. The quality of the synthesized beam is used to evaluate the image quality. We add the number of antennas, orbit, regolith properties, integration time, duty cycle, bandwidth, and observing frequency into the simulation and observe how the results of the simulated observations are affected by these factors. 

\begin{table}[h]
\caption[]{Mission Requirements}
\label{Tab:Intro}

  \begin{minipage}{15cm}
    \centering


 \begin{tabular}{ll}
  \hline\noalign{\smallskip}
Parameter &  Value                \\
  \hline\noalign{\smallskip}
Number of antennas &  $\geq 2$ \\ 
Baseline range:  &           \\
~~- Earth orbit	& $20\sim14\,000~km$	\\
~~- Lunar surface  &  $10~km$       \\
Angular resolution\footnote{10 times better than best existing surveys.}:  &        \\
~~- At $0.1\sim1~MHz$	&$\sim 6^\circ$	\\
~~- At $1\sim10~MHz$  & $\sim 0.2^\circ$        \\
Beam DM (See Section~\ref{sect:22})  & $>$10        \\
Observing frequency  & $0.3\sim30~MHz$        \\
Antenna type  & Dipole        \\
Number of poloarizations  & 2        \\
  \noalign{\smallskip}\hline
\end{tabular}

   \vspace{-0.2\skip\footins}
   \renewcommand{\footnoterule}{}
  \end{minipage}
\end{table}

As a first iteration of a trade-off study between scientific requirements and engineering constraints, in this paper we set the requirements for a hypothetical space ULW mission to the values given in Table~\ref{Tab:Intro}. In Section ~\ref{sect:Esim} and~\ref{sect:Lsim}, we present simulations of an antenna array in the Earth orbit and on the lunar surface, respectively, and estimate data transmission requirements for both cases; in Section~\ref{sect:Sensi}, we discuss sensitivities; finally, in Section~\ref{sect:Sum}, we give a summary.


\section{Simulation for an Array in the Earth Orbit}
\label{sect:Esim}

\subsection{Simulation Parameters}
Compared with ground-based antenna arrays, in the Earth orbit an array experiences real-time 3D movements, and the $(u,v)$-coverage may be significantly different and vary greatly when the synthesized beam is pointing to different directions. In addition, during observations, one or more antennas may be at the backside of the Earth, where the signal from some directions is blocked. In general, a strategy to maximize the number of available antennas at certain observation periods is needed to improve the $(u,v)$-coverage, and therefore the image quality. Below, we examine a relatively simple case in which the source is never eclipsed by the Earth. The simulation parameters are listed in Table~\ref{Tab:Esim}. All three antennas are sent into the $500~km$ orbit first by one launch. Antenna 2 needs an orbital maneuver and antenna 3 needs two steps of orbital maneuvers to reach the expected orbits.

\begin{table}[h]
\begin{center}
\caption[]{Array Parameters for the Earth orbit Simulation}
\label{Tab:Esim}

 \begin{tabular}{llll}
  \hline\noalign{\smallskip}
Parameter &  Antenna 1 & Antenna 2 &       Antenna 3          \\
  \hline\noalign{\smallskip}
Perigee  & $500~km$  & $500~km$ & $700~km$\\ 
Apogee  &  $500~km$   & $700~km$  & $700~km$      \\
$i$	& $0^\circ$	& $0^\circ$ &  $0^\circ$\\
$\omega$	& $0^\circ$	& $0^\circ$ & $0^\circ$\\
$\Omega$	&\dots	& \dots& \dots \\
$t_{0}$  & 0 & 0 & 0        \\
Duty cycle  & \multicolumn{3}{c}{{$1~s$ per $90~s$ (See section \ref{sect:dc}) }}      \\
Observing frequency &  \multicolumn{3}{c}{$10~MHz$}        \\
Phase center  & \multicolumn{3}{c}{North Celestial Pole }      \\
  \noalign{\smallskip}\hline
\end{tabular}

\end{center}
\end{table}

\subsection{Simulation Results}
\label{sect:22}


\begin{figure}[h]
  \centering
  \includegraphics[width=4in]{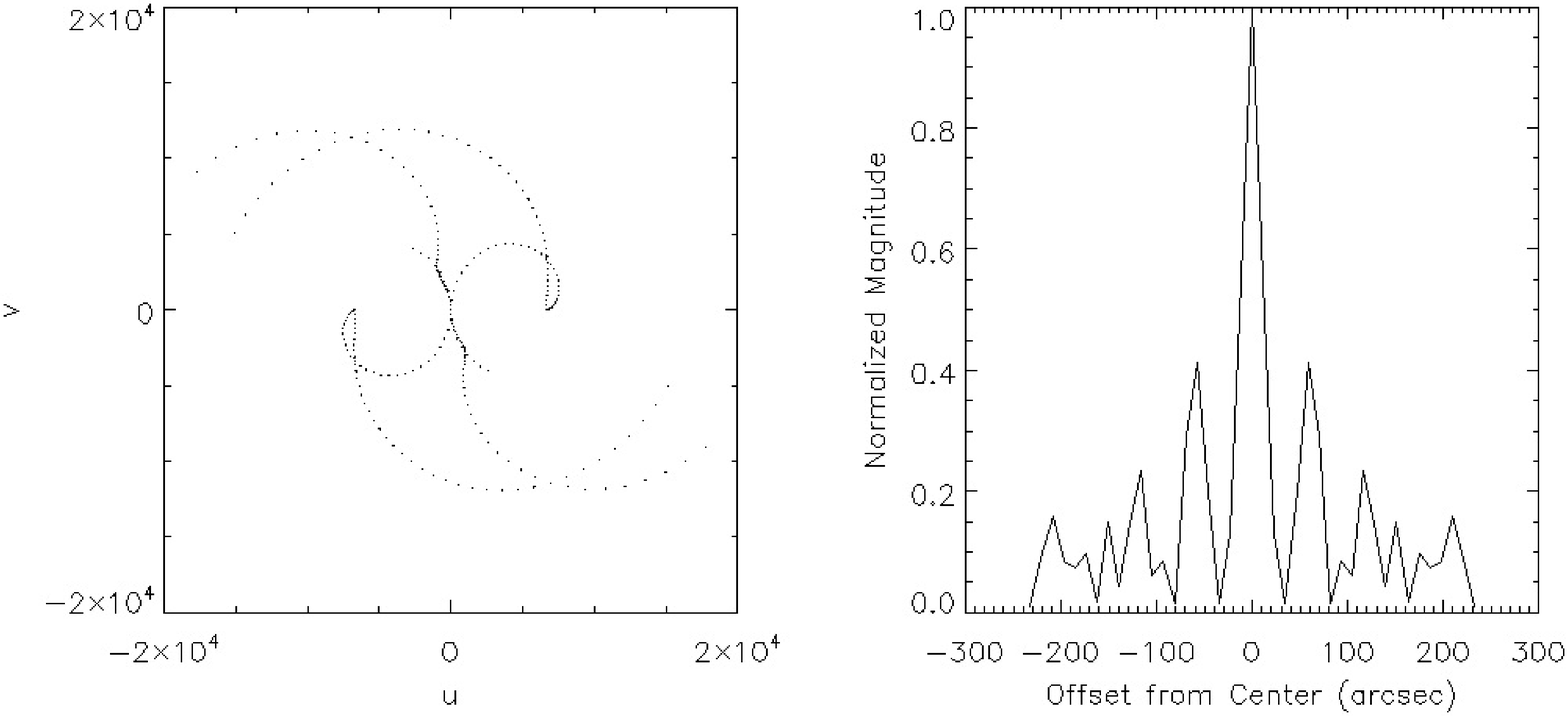}
  \caption{$(u,v)$-coverage and beam profile for a 1-h observation}
\label{Fig:1}
\end{figure}

\begin{figure}[h]
  \centering
  \includegraphics[width=4in]{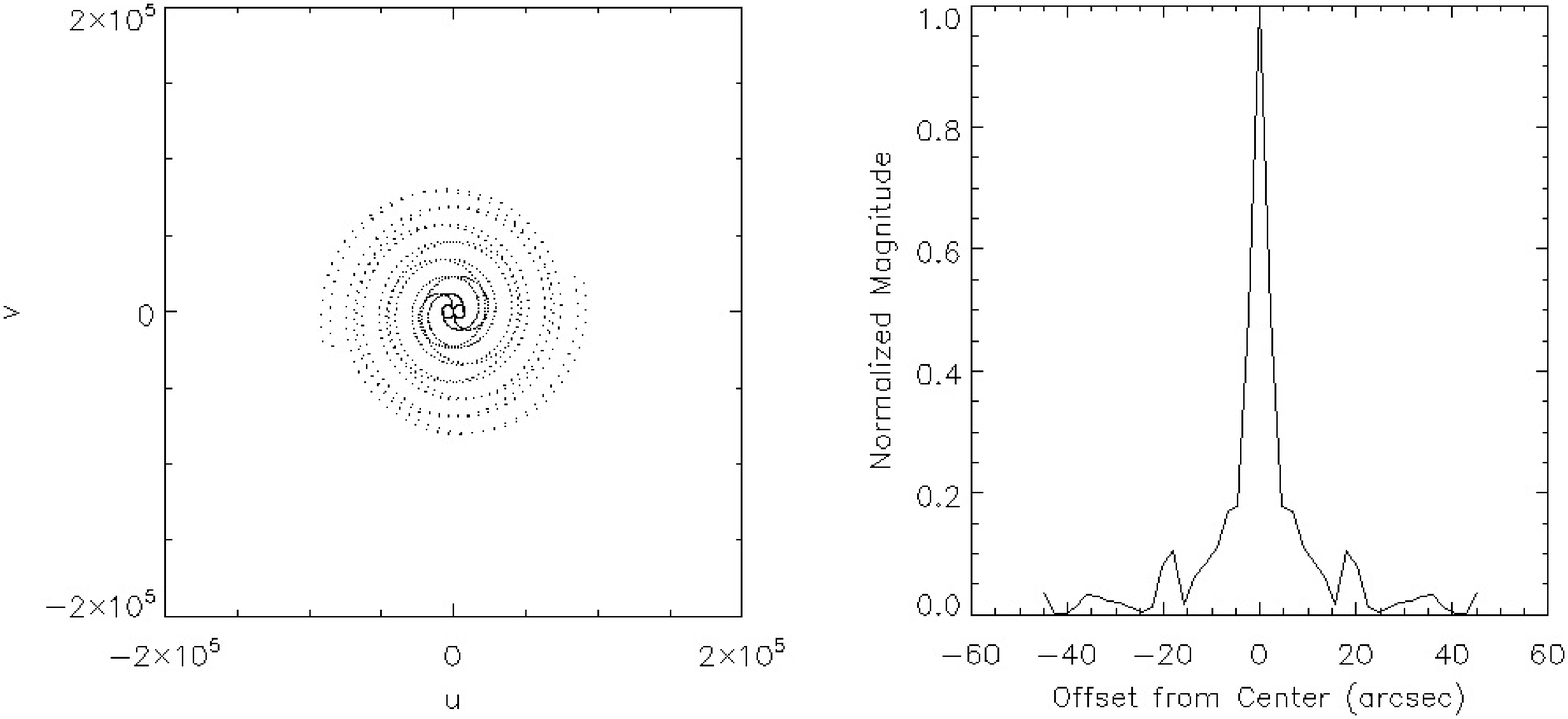}
  \caption{Same as Fig. 1, but for 6-h}
\label{Fig:2}
\end{figure}

\begin{figure}[h]
  \centering
  \includegraphics[width=4in]{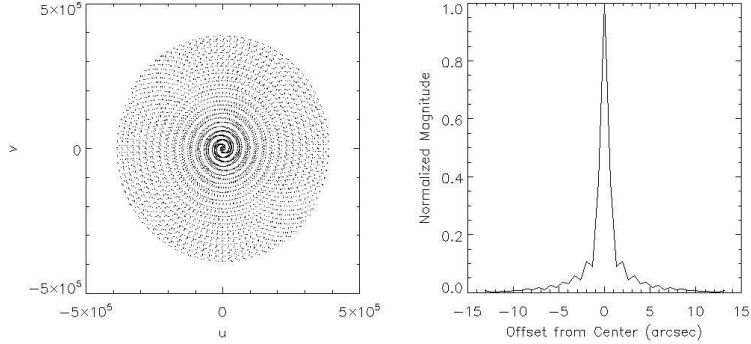}
  \caption{Same as Fig. 1, but for 24-h}
\label{Fig:3}
\end{figure}

Figures~\ref{Fig:1},~\ref{Fig:2} and~\ref{Fig:3} show the $(u,v)$-coverages and cross-sectional views of the synthesized beams for integration times of $1~h$, $6~h$ and $24~h$, respectively. As shown in the right hand panels of these figures, the beam directivity improves with longer integrations. The diffraction-limited theoretical resolution and the beam directivity measure (DM), defined as the maximum value of the main beam divided by the RMS ``noise'' in the beam map, are shown in Table~\ref{Tab:dm} for different observation periods. Observation quality for sources at other directions can be simulated using the same method, with the effect of the Earth eclipses considered if necessary.

\begin{table}[h]
\begin{center}
\caption[]{Beam DM for Different Observation periods}
\label{Tab:dm}

 \begin{tabular}{lll}
  \hline\noalign{\smallskip}
Observation period ($h$) & Theoretical resolution ($^{\prime\prime}$) & Beam DM           \\
  \hline\noalign{\smallskip}
1	& 12	& 14 \\
6  & 2	&31\\
24  & 0.7	& 60\\
  \noalign{\smallskip}\hline
\end{tabular}
\end{center}
\end{table}

\subsection{Data Transmission Requirements}
\label{sect:dc}
We assume that the ULW signal received by the antenna is sampled without down-conversion, and all the signals are transmitted to the ground for processing. The data volume is the product of the bits per sample, sampling frequency, integration time, and number of polarizations. For an observing frequency of $10~MHz$, the sampling frequency must be at least $20~MHz$ for lossless recovery, according to the Nyquist sampling theorem. We also assume that each sample occupies 1 bit of storage per polarization and that there are two polarizations per antenna, in which case each antenna will achieve a data rate of $10~MHz\times 2~\times (1~/~8)~byte ~per~ sample\times 2~polarizations=40~Mbps$, not including housekeeping and auxiliary data. 

We adopt the parameters of a recent study by CNES (Peragin et al.~\cite{Peragin12}) to evaluate constraints on downlink requirements in X-band. This study shows that a volume of $39.9~Gb$ data can be downloaded in X band with a $5~m$ station, considering a $95~min$ orbit with $8~min$ pass window on average and 3 passes per day. An observation duty cycle is applied to reduce the data volume because the downlink speed is not fast enough if data are taken all the time. As shown in Table~\ref{Tab:data}, if the observation duty cycle is set to 1 second per 90 seconds, the bandwidth requirement is $\sim27~Mbps$ per antenna station, which is acceptable.

\begin{table}[h]
\begin{center}
\caption[]{Data Transmision Requirements for the Earth orbit array}
\label{Tab:data}

 \begin{tabular}{ll}
  \hline\noalign{\smallskip}
Parameter & Value        \\
  \hline\noalign{\smallskip}
Observing frequency	& $10~MHz$ \\
Sampling frequency	& $20~MHz$ \\
Duty cycle  & $1~s$ per $90~s$ \\
Bits per sample  & 1 \\
Number of polarizations  & 2 \\
Data rate requirement~(one station)  & $27~Mbps$ \\
  \noalign{\smallskip}\hline
\end{tabular}
\end{center}
\end{table}

\subsection{Discussion}
In our simulations, we assume that the antennas are nearly collocated when the observation starts. This is not generally true. The $(u,v)$-coverage for an arbitrary initial configuration or an arbitrary period of time, and the statistical properties thereof, can be studied based on the method given in this section.

In the design of some space VLBI missions, the antennas are always parallel to the surface of the Earth. With multiple antennas in different orbit positions this design will result in unparalleled beams, which in general will reduce sensitivity and $(u,v)$-coverage. For our simulations above, however, this issue does not exist because the orbit plane is on the equator and the antennas always point to the celestial poles that is the phase center. In a simulation of a general case the issue with unparalleled antennas should be considered.
%

If one needs to make high time resolution observations, such as of solar radio bursts, the $(u,v)$-coverage of each snapshot may be very poor, especially when using a small number of antennas. For such cases, further investigation is required to develop mission configurations and data reduction techniques.

\section{Simulation for an Array on the Lunar Surface}
\label{sect:Lsim}
\subsection{Lunar Regolith Modeling}
The relative permittivity of the lunar regolith is approximately 6 (Woan~\cite{Woan96}). For a dipole lying on the lunar surface, the response to the signal transmitted from below the lunar surface will be 4.5 times that for a signal from above (Woan~\cite{Woan96}).

\begin{figure}[h]
  \centering
  \includegraphics[width=3in]{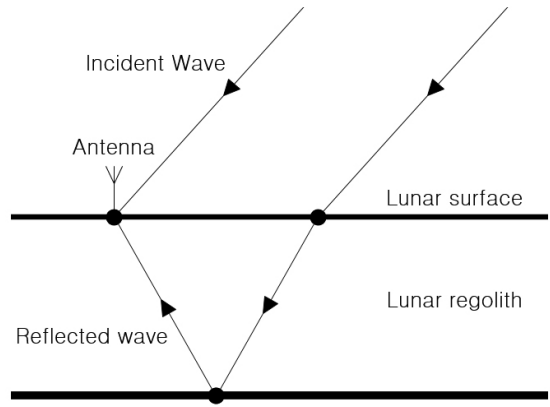}
  \caption{Simplified signal transmission path model for the lunar regolith}
\label{Fig:4}
\end{figure}

The depth of the lunar regolith is about $10~m$ (Lindsay~\cite{L76}). If the observing frequency is $20~MHz$, the signal transmission distance in the regolith is about 2 times the wavelength, with a travel time of about $100~ns$. The signal an antenna receives is the sum of the signals from the sky and from beneath the lunar surface. For the same source, the two signal components are emitted at different times due to reflection of the signal in the lunar regolith. Greater incidence angles result in a longer optical path in the regolith. The regolith will affect the coherence of signals and therefore the imaging quality in complex ways. In the following, we quantitatively show the effect with a simplified model to explain how the lunar regolith affects such observations.

Figure~\ref{Fig:4} shows a simplified model of the lunar regolith. Below the regolith is bedrock, which serves as a reflector. This model ignores the complex structure of the lunar regolith and bedrock; it assumes that the depth of the regolith is constant, the regolith is uniform and isotropic, and the boundary between regolith and bedrock is flat.

\subsection{Simulation Results}



\begin{table}[h]
\caption[]{Array Parameters for the Lunar Surface Simulation}
\label{Tab:lsim}
  \begin{minipage}{15cm}
    \centering
\begin{tabular}{ll}
  \hline\noalign{\smallskip}
Parameter &  Value                \\
  \hline\noalign{\smallskip}
Number of antennas & 2 \\
Antenna latitude & $45^\circ N$ \\
Antenna configuration & East-west, $10~km$ seperation \\
Hour-angle sampling range & $-60^\circ\sim60^\circ$ \\
Observation wavelength & $15~m$ \\
Source position & North Celestial Pole \\
Lunar regolith: \\
~~- Depth  & $10~m$ \\
~~- Refractive index (Lindsay~\cite{L76}) & 1.5 \\
~~- Signal attenuation ratio\footnote{Defined as the ratio of the signal magnitude from beneath the lunar surface to that from above, without considering antenna gain.} & 0.1 \\
Antenna type & Short dipole \\
Number of samples & 100 \\
  \noalign{\smallskip}\hline
\end{tabular}
   \vspace{-0.2\skip\footins}
   \renewcommand{\footnoterule}{}
  \end{minipage}

\end{table}

\begin{figure}[h]
  \centering
  \includegraphics[width=3in]{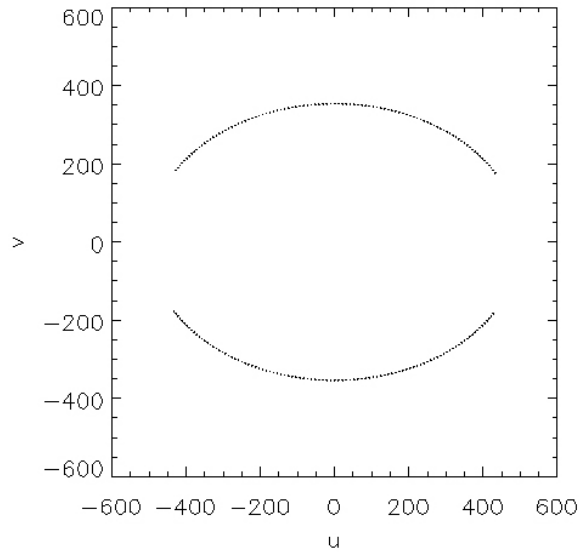}
  \caption{$(u,v)$-coverage of a two-antenna array}
\label{Fig:5}
\end{figure}

\begin{figure}[h]
  \centering
  \includegraphics[width=4in]{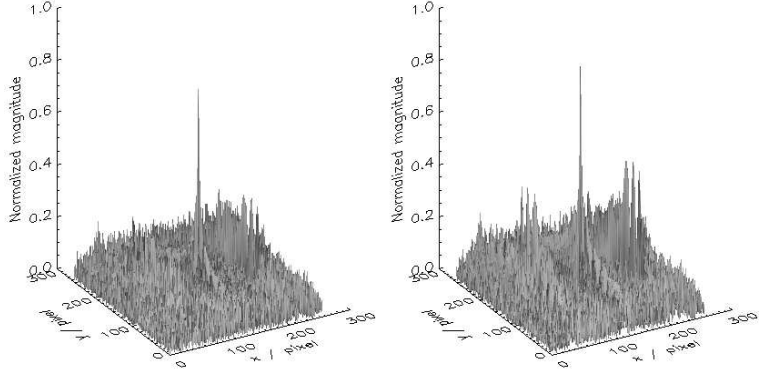}
  \caption{Normalized beam profile when ignoring the effect of the lunar regolith (left) and when taking it into account (right)}
\label{Fig:6}
\end{figure}

Table~\ref{Tab:lsim} shows the parameters we used in the lunar surface simulation, considering there are only two antennas, one on the lander and one on the rover. Both of them are on the near side of the moon. Figures~\ref{Fig:5} and~\ref{Fig:6} show the $(u,v)$-coverage and beam profiles of the simulation. By comparing the two beam profiles in Figure~\ref{Fig:6}, one can see that the beam DM is smaller when the effect of the lunar regolith is taken into account (Table~\ref{Tab:ldm}). We conclude that the lunar regolith will have an undesirable effect on imaging, which confirms the above discussion.

\begin{table}[h]
\begin{center}
\caption[]{Beam DM for Different Parameters}
\label{Tab:ldm}
\begin{tabular}{ll}
  \hline\noalign{\smallskip}
Lunar regolith &  Beam DM                \\
  \hline\noalign{\smallskip}
Not considered & 17 \\
Taken into account & 15 \\
  \noalign{\smallskip}\hline
\end{tabular}
\end{center}
\end{table}

\subsection{Data Transmission Requirements}

Compared with the Earth orbit antenna array, the data transmission bandwidth for a lunar array is much more limited, approximately $2.5~Mbps$ in the case of the Chang'E-1 mission. We assume the antennas are located on the near side of the moon and are visible from the ground station 8 hours a day on average. Further assume that each antenna has a dedicated downlink channel of $2.5~Mbps$ capacity, downlinking data 8 hours per day, for an observing frequency of $10~MHz$, 1-bit sampling and 2 polarizations, a duty cycle of $1~s$ per $48~s$ is needed to meet the data transmission bandwidth requirement. According to long-term plan of the Chang'E Program, we assume the lander and the rover to be on the near side of the moon. If they are on the far side of the moon, a relay satellite orbiting around the moon is needed. The transmission capability of the relay satellite is assumed to be the same as the lander and the rover, so the calculations are still applicable.

\section{Array Sensitivity}
\label{sect:Sensi}
We assume basic observational parameters to derive the flux sensitivity and brightness temperature sensitivity of the antenna arrays. When the antenna beam is much larger than the angular size of the source, one is concerned with the flux sensitivity; when the beam is comparable to or smaller than the source, one is concerned with brightness temperature sensitivity.

The flux sensitivity can be defined as
\begin{equation}
\label{eq:1}
   \sigma_s = \frac{2k{T_{sys}}}{A_e\sqrt{2n(n-1)\Delta\nu\tau{n_p}}}(Jy),
\end{equation}
where $T_{sys}$ is the antenna system temperature, $n$ is the number of antennas, $\Delta\nu$ is the bandwidth or channel width, $\tau$ is the integration time, $n_{p}$ is the number of polarizations, $A_{e}$ is the effective collecting area of one dipole, which is defined as $3\lambda^{2} / 8 \pi $ (Woan~\cite{Woan96}), and $k$ is the Boltzmann constant ($1380~Jy~K^{-1}~m^{2}$).

The brightness temperature sensitivity is defined as
\begin{equation}
    \label{eq:2}
{
 \sigma_t = \frac{D^2{T_{sys}}}{A_e\sqrt{2n(n-1)\Delta\nu\tau{n_p}}}(K),
}
 \end{equation}
where $D$ is the maximum baseline of the interferometer array.

\begin{table}[h]
\begin{center}
\caption[]{Flux Sensitivity for Different Array Parameters}
\label{Tab:flux}
 \begin{tabular}{lllllll}
  \hline\noalign{\smallskip}
 Number of  &  \multicolumn{2}{c}{$\lambda~=~$$10~m$} &\multicolumn{2}{c}{ $\lambda~=~$$30~m$} & \multicolumn{2}{c}{$\lambda~=~$$300~m$ }  \\
antennas & \multicolumn{2}{c}{$(30~MHz)$} &\multicolumn{2}{c}{ $(10~MHz)$ }&\multicolumn{2}{c}{ $(1~MHz)$ }   \\
\cline{2-7}
 &Bandwidth & Sensitivity &Bandwidth & Sensitivity &Bandwidth & Sensitivity \\
 &($MHz$) & ($Jy$) &($MHz$) &($Jy$)  &($MHz$) &($Jy$)  \\
  \hline\noalign{\smallskip}
2	& 	 & 7  &  & 13&   & 31 \\
3	& 1	 & 4  & 1 & 8& 0.1  & 18 \\
  10      &  &1   &  &2 &  & 5  \\
  \hline\noalign{\smallskip}
2	& & 2  &  & 4&  & 10 \\
 3      &  10 & 1  & 10 &2 &  1 & 6 \\
  10       & & 0.4  &  & 0.6 &   & 1 \\
  \noalign{\smallskip}\hline
\end{tabular}
\end{center}
\end{table}

\begin{table}[h]
\begin{center}
\caption[]{Temperature Sensitivity for Different Array Parameters}
\label{Tab:temp}
  \begin{minipage}{15cm}
    \centering
 \begin{tabular}{llll}
  \hline\noalign{\smallskip}
Bandwidth ($MHz$) & Number of antennas & \multicolumn{2}{c}{Temperature sensitivity ($\sigma_t$) at $\lambda = $$30~m~(K)$}     \\
\cline{3-4}
 & & Earth orbit: $D~=~14\,000~km$\footnote{The maximum baseline in this orbit is used for calculation. A shorter baseline results in better temperature sensitivity. See section \ref{sect:Sensi}.} & Lunar surface: $D~=~10~km$  \\
  \hline\noalign{\smallskip}
  & 2 &$1 \times 10^{12}$  &  $5 \times 10^{5}$\\
 1 & 3 & $6 \times 10^{11}$  & $2 \times 10^{5}$ \\
  & 10 & $1 \times 10^{11}$ & $5 \times 10^{4}$  \\

	& 2	& $3 \times 10^{11}$ &$2 \times 10^{5}$ \\
10	& 3	& $2 \times 10^{11}$ & $6 \times 10^{4}$ \\
   & 10 & $4 \times 10^{10}$ & $2 \times 10^{4}$  \\

  \noalign{\smallskip}\hline
\end{tabular}

   \vspace{-0.2\skip\footins}
   \renewcommand{\footnoterule}{}
  \end{minipage}
\end{center}
\end{table}

Tables~\ref{Tab:flux} and \ref{Tab:temp} show the two types of sensitivity for different observation parameters (bandwidth, observation wavelength, and antenna number, assuming $n_{p} = 2$, $\tau = $$86\,400~s$. A reduction of $36\%$ of the sensitivity caused by 1-bit sampling has been taken into account (Thompson et al.~\cite{TMS2001}). For ULW observations, $T_{sys}$ is dominated by the Galactic background radio emission, at frequencies above $2~MHz$,  it can be approximated by a power law (Jester \& Falcke~\cite{JF09}): 

\begin{equation}
    \label{eq:3}
{
	T_{sys} = 16.3\times 10^6K(\frac{\nu}{2MHz})^{-2.53}
},
 \end{equation}

at frequencies below $2~MHz$, we use the value predicted by the RAE-2 spacecraft, $2 \times 10^{7}~K$ at $1~MHz$ (Novaco \& Brown~\cite{NB78}).

We can then choose array parameters that meet the requirements to observe specific radio sources at a certain level of flux density based on these results. For example, the average flux density of 3C 273 is about $10^{2}~Jy$ at $30~MHz$, according to the flux densities of active radio sources in the $10~kHz\sim100~MHz$ range from the ESA~(\cite{BLV97}) design study, adapted from Zarka et al.~(\cite{ZQR97}).

For thermal sources, for example regions of ionized gas surrounding massive stars, the maximum brightness temperature is approximately $2 \times 10^{4}~K$. Optically-thick thermal bremsstrahlung emission dominating the quiet Sun produces brightness temperatures of order $10^{6}~K$ (White~\cite{White07}). For non-thermal sources, the brightness temperature of the Galactic radio background radiation from the polar regions is $2.3 \times 10^{7}~K$ at $1~MHz$ (Cane~\cite{Cane79}), and can be as high as $10^{12}~K$ for synchrotron radiation sources (Wilson et al.~\cite{Wilson2009}). 

We also note that temperature sensitivity must be high enough to achieve sufficient SNR (Signal to Noise Ratio) :
\begin{equation}
    \label{eq:3}
{
SNR=\frac{T_{sys}}{\sigma_t}=\frac{A_e}{D^2}\sqrt{n(n-1)\Delta\nu\tau}=f\sqrt{\Delta\nu\tau}\frac{\sqrt{n(n-1)}}{n}},
 \end{equation}
where $f = nA_{e}/D^{2}$ is the filling factor of the interferometer. An observation with three antennas, 90 days observation period, $1~s$ per $90~s$ duty cycle and $1~MHz$ bandwidth will achieve SNR of 20.

The maximum baseline is the key factor when considering trade-offs between sensitivity and angular resolution. Taking a low Earth orbit array as an example, an observation with 10 antennas, $10~MHz$ observing frequency, $10~MHz$ bandwidth, 90 days observation period, $1~s$ per $90~s$ ducy cycle and $14\,000~km$ maximum baseline, will achieve a temperature sensitivity of $4 \times 10^{10}~K$. In such case, the angular resolution is about $0.5\arcsec$, far from being able o resolve the synchrotron self-absorbed sources with $10^{12}~K$. Because noise in brightness temperature ($\sigma_t$) is proportional to $D^{-2}$ and angular resolution is proportional to $D^{-1}$, reducing maximum baseline will allow various kind of sources to be observable with sufficient temperature sensitivity and angular resolution, such as the Galactic radio background radiation at arcminute resolution and solar bursts. There are multiple methods to limit maximum baseline for an Earth orbit array, either by choosing observing windows so that data for extended sources are only taken when the antennas are close to each other, or by formation flight.

\section{Summary}
\label{sect:Sum}

We demonstrate a simulation framework for ULW space arrays in the Earth orbit and on the lunar surface to help the overall system design of future ULW missions. We present simulations, estimate the data transmission requirements and calculate sensitivities for both cases. For the lunar surface array case, we create a simplified regolith model, the results show that the regolith would have an undesirable effect on the observation. For the Earth orbit array case, we discuss the relation of the maximum array baseline to the angular resolution and the brightness temperature sensitivity to show the trade-off between brightness temperature sensitivity and angular resolution.

\label{lastpage}
\end{document}